\documentclass[twocolumn,amsmath,amssymb]{revtex4}
\usepackage{graphicx}
\usepackage{setspace}
\usepackage{epsfig}
\begin{document}
\title{Velocity of front propagation in the 
       epidemic model $A+B\rightarrow2A$.}
\author{ Niraj Kumar and Goutam Tripathy }
\address{Institute of Physics, Sachivalaya Marg, Bhubaneswar 751005,
India} 
\begin{abstract}
    We study front propagation in the irreversible epidemic model 
     $A+B\rightarrow2A$ in one dimension 
    . Here, we allow the particles $A$ and $B$ to diffuse 
    with rates $D_A$ and $D_B$, which, in general, may be 
    different. We find analytic estimates for 
    the front velocity by writing truncated master equation
    in a frame moving with the rightmost $A$ particle. The
    results obtained are in reasonable agreement with the simulation 
    results and are amenable to systematic improvement. We also 
    observe a crossover from the linear dependence of front velocity $V$
    on $D_A$ for smaller values of  $D_A$ to $V\propto \sqrt{D_A}$ for
    larger $D_A$, but numerically still significantly different from
    the mean field value. The deviations reflect the role of internal 
    fluctuations which is neglected in the mean field description.
\end{abstract}
\maketitle
\section{Introduction}
In many natural phenomena we encounter propagating fronts separating
different phases \cite{sar}. In this paper we study the 
autocatalytic reaction
$A+B\rightarrow2A$ in a one dimensional lattice. Models of this type 
have been studied actively because of several interesting
features of the process \cite{mai1,mai2,mai3,sander}. We
can think of this 
reaction as a model for various spreading phenomena such as an 
epidemic: particle $A$ is an infectious agent which infects a healthy
particle $B$ on contact. We are interested in finding the velocity of 
propgagation of this infection. The macroscopic description for this
process is given by the following mean field equation \cite{fisher}.
\begin{eqnarray}{\label{e1}}
\frac{\partial\rho_A}{\partial t}=D_A\triangle\rho_A
   +\epsilon\rho_A\rho_B
 =D_A\triangle\rho_A+\epsilon\rho_A(\rho_0-\rho_A)
\end{eqnarray}
Here, $\rho_{A}(x,t)$ is the local density of $A$ particle at position
$x$ and time $t$, $D_A$ is its diffusion coefficient and $\epsilon$ 
is the reaction rate of the particle. In the second line we 
have used  $\rho_A+\rho_B=\rho_0$ which is constant. Equation(
\ref{e1}) arises in the macroscopic description of diverse physical
processes and serves as a generic model 
for front propagation in a system in which a stable state invades an
unstable state \cite{sar}. Equation (\ref{e1}) allows a family of
travelling wave solutions with speed $V\ge 2\sqrt{\epsilon\rho_0 D_A}$
invading the unstable state $\rho_A=0$ from the stable state 
$\rho_A=\rho_0$. However, for steep enough initial conditions, the
selected asymptotic speed is the minimum speed $V_0=2\sqrt
{\epsilon\rho_0 D_A}$ \cite{sar}. In the microscopic lattice model 
in low dimensions, the discreteness effects alter the dynamical 
properties of the propagating  front and 
the results obtained are different from that using deterministic
mean field equation \cite{derrida}\cite{kessler}. Here, in this paper 
we will present the Monte Carlo simulation results and approximate analytic 
prediction for the front velocity in a one dimensional lattice 
model for $A+B\rightarrow2A$ defined in the next section (II).
\section{Model and front velocity}
In our simulation, we start with a 1D lattice $[0,\infty)$
initially randomly filled with $B$ particles with overall concentration
$\rho_0$. We have taken hard core exclusion
into account: any site can be occupied by maximum one 
particle. At the left end of the lattice we place an $A$ particle. We 
allow both the particles $A$ and $B$ to diffuse to the nearest
neighbour empty sites with rates $D_A$ and 
$D_B$ respectively. When a 'sick' particle $A$  encounters the 'healthy'
one $B$ sitting on a neighbouring site, then $B$ gets infected with rate 
$\epsilon$ and is converted to an $A$ particle. In this way the front
( rightmost $A$ ) propagates and we wish to find the velocity of this 
front.

\begin{figure}
 \centering
 \includegraphics[bb=90 144 269 277]{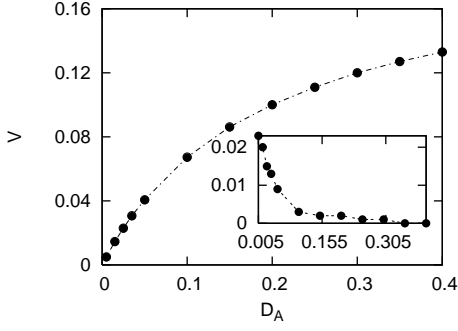}
 \caption{Front velocity $V$ as a function of $D_A$ for
      $D_B=0.05, \epsilon=0.20, \rho_0=0.5$. The broken
      line represents the front velocity obtained 
      by simulation while the filled circles corresponds 
      to the velocity obtained by using Eq. (\ref{e2}) and
      substituting the values of $\rho_1^+$ and $\rho_1^-$
      from simulation. Inset: The differenece $|\rho_1^--\rho_0|$  
      as a function of $D_A$, $\rho_1^-$ taken directly 
      from simulation.}
 \label{fig:vel_1}
 \end{figure}

\begin{figure}
 \centering
 \includegraphics{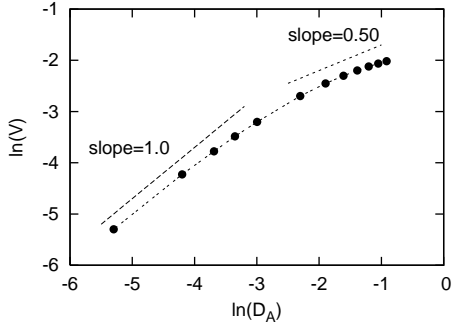}
 \caption{Log-log plot for the front velocity as a function of $D_A$ for
 $D_B=0.05, \epsilon=0.20,\rho_0=0.5$. Here $V\sim D_A^{\alpha}$ and we 
 notice a crossover from $\alpha\approx 1.0$ to $\alpha\approx 0.50$.}
 \label{fig:vel_lnln}
 \end{figure}

In Ref \cite{mai1}, authors studied a similar problem using Smouluchowski
approach \cite{smol} and obtained the steady state density for
$B$ and $A$ particles at a distance $x$ from the moving front 
particle (at $x=0$)  
as: $\rho_B(x>0)=\rho_0(1-e^{-Vx/(D_A+D_B)})$ and $\rho_A(x<0)=\rho_A$
respectively. By 
invoking the equality of particle fluxes to the left($j_-=V\rho_A$)
and right[($j_+=(D_A+D_B)(\partial/\partial x)\rho_B(+0))$] they 
found $\rho_A=\rho_0$, but it does not fix the front propagation 
velocity $V$. In the present paper, we try to find the 
expression for the front velocity by visualising the 
front as random walker moving with forward and backward rates as $P_+$ 
and $P_-$ respectively, and thus speed of the front $V=P_+-P_-$. 
In order to compute $P_+$,  we notice that there 
are two ways in which the front will move to the right: (1) when 
the front diffuses to the right neighbouring site, provided it
is empty with rate $D_A(1-\rho_1^+)$, here $\rho_1^+$ is the 
probability that the site just ahead of the front is occupied, (2) when
the site just ahead of front is occupied by $B$ particle, then due to 
reaction it 
changes the neighbouring $B$ into $A$ and thus the front moves forward 
with rate $\epsilon\rho_1^+$ and hence $P_+=\epsilon\rho_1^+ 
+ D_A(1-\rho_1^+)$. Similarly, the front moves backward 
due to its diffusive move to the left empty neighbouring site and 
hence $P_-=D_A(1-\rho_1^-)$, $\rho_1^-$ is the probability
that site just behind the front is occupied. Using these values for
 $P_+$ and $P_-$, we have for the asymptotic velocity,
 \begin{eqnarray}{\label{e2}}
   V=\epsilon\rho_1^++D_A(\rho_1^{-}-\rho_1^{+}),
 \end{eqnarray}
 which is exact, given $\rho_1^+$ and $\rho_1^-$. The multi-particle
 nature of the problem is embodied in finding $\rho_1^+$ and $\rho_1^-$.
 The fact that equation (\ref{e2}) is exact can be seen from the
 excellent agreement between simulation result of $V$ and its 
 estimate using values of $\rho_1^+$ and $\rho_1^-$ obtained directly
 from simulation, Fig.(\ref{fig:vel_1}). Here, we also observe a crossover
 from linear dependence of $V$ on $D_A$ for small $D_A$ to  
 $V\propto\sqrt{D_A}$ for large $D_A$ as shown in Fig. (\ref{fig:vel_lnln}),
 although we have verified that $V\ne V_0$(mean field value). 
 
\begin{figure}
 \centering
 \includegraphics[width=3.0in,height=2.0in]{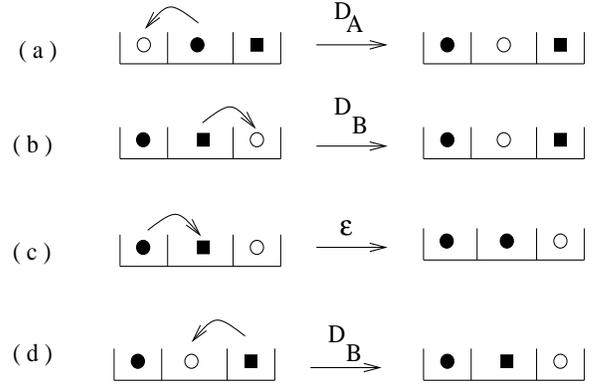}
 \caption{Microscopic moves: Filled circles/squares represent 
     $A/B$ particles. (a) Diffusion of $A$ to the left empty site
     leading to transition from the state $(AB)$ to $(A\phi)$, (b)
     Diffusive jump of $B$ particle to its right empty site leading
     to transition from $(AB)$ to $(A\phi)$, (c) Transition 
     from $(AB)$ to $(A\phi)$ due to reaction and (d) Diffusion of $B$
     to its left and thus making transition from $(A\phi)$ to $(AB)$.}
 \label{fig:moves}
 \end{figure}

 From \cite{mai1}, the density profile of $A$ particles behind the
 front is constant and equal to $\rho_0$. Thus we can make approximation
 $\rho_1^-=\rho_0$. But we note that this approximation fails to work
 when $D_A=0$ where we get $\rho_A=1$ at all the sites behind the front
 irrespective of $\rho_0$. Thus the approximation $\rho_1^-=\rho_0$ works
 better for larger values of $D_A$ while for smaller $D_A$ it appears to
 be a poor approximation as shown in the Fig. (\ref{fig:vel_1}, inset).
 Now using the approximation $\rho_1^-=\rho_0$  we try to find  
 analytic expression for $\rho_1^+$. In the following we find an 
 approximate analytic expression for $\rho_1^+$ using the technique 
 developed in \cite{ng}. The technique involves writing a truncated 
 master equation in a frame moving with the front ( i. e. the 
 rightmost $A$ particle). For example, the simplest one is the set of
 two states $\{A\phi, AB\}$. In this truncated representation each 
 state contains the occupancy at two sites ($l=2$), the leftmost $A$ 
 represents the front particle while $\phi$ and $B$ stand for empty 
 and a site occupied by $B$ just ahead of the front respectively. These
 two states make transitions between each other due to reaction or 
 diffusion of particles, see Fig. (\ref{fig:moves}). For example, in 
 the transition shown in Fig. (\ref{fig:moves}a), the configuration 
 $AB$ changes to $A\phi$ if $A$ particle diffuses to its left empty 
 site and it takes place with rate $D_A(1-\rho_1^-)$.
 In Fig. (\ref{fig:moves}b), diffusion of $B$ particle in the 
 configuration $AB$ to its right neighbouring empty site changes 
 $AB$ to $A\phi$. If the probability of occupancy of $B$ particle 
 at a {\it{second}} site ahead of the front is denoted by $\rho_2^+$,
 then $AB\rightarrow A\phi$
 with rate $D_B(1-\rho_2^+)$. In the Fig. (\ref{fig:moves}c), when 
 $A$ infects
 $B$ in the configuration $AB$ then $AB\rightarrow A\phi$ if the 
 second site ahead of the front is empty and this occurs with 
 rate $\epsilon(1-\rho_2^+)$. Similarly, in Fig(\ref{fig:moves}d), 
 $A\phi$ changes to $AB$  with rate $D_B\rho_2^+$. Thus
 considering all the transitions between the two states $A\phi$ and $AB$
 and approximating $\rho_1^-=\rho_0$, the following master equation 
 may be written \cite{kampen}:
 \begin{eqnarray}{\label{e3}}
 \frac{dp_{A\phi}}{dt} &=& \{D_B(1-\rho_2^+)+D_A(1-\rho_0)+\epsilon(1
  -\rho_2^+)\}p_{AB}\nonumber\\&-&(D_A+D_B)\rho_2^+p_{A\phi},\nonumber\\
 \frac{dp_{AB}}{dt} &=& (D_A+D_B)\rho_2^+p_{A\phi}-\{D_B(1-\rho_2^+)
            +D_A(1-\rho_0)\nonumber\\&+&\epsilon(1-\rho_2^+)\}p_{AB}.
 \end{eqnarray} 
 In the steady state, by definition, $\rho_1^+=p_{AB}$ and thus from
 Eq.(\ref{e3}) with the normalization $p_{A\phi}+p_{AB}=1$ we obtain
 \begin{eqnarray}{\label{e4}}
 \rho_1^+=\frac{(D_A+D_B)\rho_2^+}{\{D_B+D_A\rho_2^++D_A(1-\rho_0)
       +\epsilon(1-\rho_2^+)\}}
 \end{eqnarray}
 
\begin{figure}
 \centering
 \includegraphics{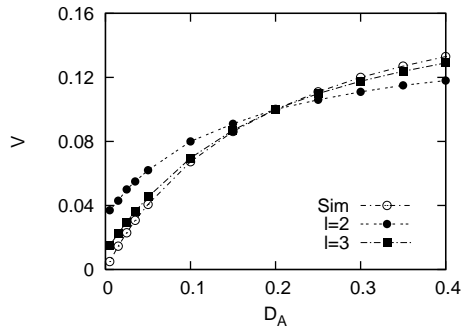}
 \caption{Velocity V as a function of $D_A$ for $D_B=0.05, \epsilon
       =0.20,\rho_0=0.5$. Filled circle($\bullet$) represents 
       analytic estimate for $l=2$, that is, for states 
       $\{A\phi, AB \}$. Filled square corresponds to the results 
       obtained for $l=3$, that is, for the 
       states: $\{A\phi\phi, A\phi B, AB\phi, ABB \}$ while the open circle
       ($\circ$) is the simulation results for the front 
       velocity. Here, we note that as $l$ increases the analytic
       prediction curve comes closer to the simulation results.}
 \label{fig:vel_2}		
 \end{figure}

\begin{figure}
  \centering
  \includegraphics{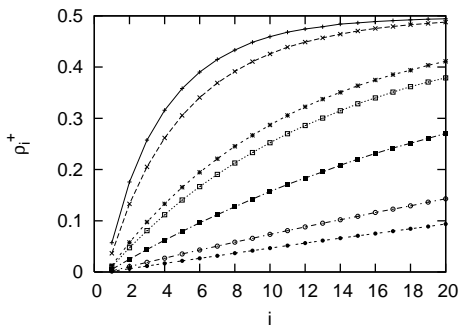}
  \caption{Density profile ahead of the front particle as a 
  function of $D_A$ while keeping $D_B=0.05, \epsilon=0.20$, and 
  $\rho_0=0.5$ fixed. The simulation data from top to bottom
  corresponds to $ D_A=0.025, 0.015, 0.005, 0.004, 0.002, 0.0005, 0.0 $.
  Here we note that as $D_A\rightarrow 0$, $\rho_1^+\rightarrow 0$.
  }
   \label{fig:den_profile}
  \end{figure}

 Thus, in order to find $\rho_1^+$ we need to know 
 $\rho_2^+$ and as a first approximation assuming 
 $\rho_2^+=\rho_0$, we find 
 \begin{eqnarray}{\label{e5}}
 \rho_1^{+}=\frac{(D_A+D_B)\rho_0}{D_A+D_B+\epsilon(1-\rho_0)}
 \end{eqnarray}
 Using this approximation for $\rho_1^+$ in Eq. (\ref{e2}) 
 we obtain the following expression for the velocity,
 \begin{eqnarray}{\label{e6}}
 V=\frac{\epsilon\rho_0(2D_A+D_B-D_A\rho_0)}{D_A+D_B
    +\epsilon(1-\rho_0)}
 \end{eqnarray}
 We have shown the estimate for the velocity as obtained from 
 Eq. (\ref{e6}) in Fig. (\ref{fig:vel_2}). Here we note 
 that when $\epsilon=D_A$, the above approximate result for $V$ 
 satisfies the simulation result nicely. However, as the magnitude
 of $\epsilon-D_A$ increases the analytic prediction for $V$ departs
 gradually from the simulation results. From Eq. (\ref{e2}), we
 note that when $D_A=\epsilon$ the expression for $V$ is independent 
 of $\rho_1^+$ and hence we obatain nice agreement between the 
 theory and simulation results. But the dependece of $V$ on $\rho_1^+$
 increases gradually with the inceasing magnitute of 
 $\epsilon-D_A$ and hence we observe poor agreement between 
 analytic and simulation results as we move away from the point 
 $D_A=\epsilon$. In order to get better prediction 
 for the velocity, especially for larger values of 
 $\epsilon-D_A$, we need to study the states having larger 
 number of particles. For example, we studied the following set
 of states($l=3$): $\{A\phi\phi, AB\phi, A\phi B, ABB\}$. If we denote the
 probability of occupancy at {\it{third}} site ahead of the front 
 as $\rho_3^+$, the
 evolution of these states is described by the following master equation.
 \begin{eqnarray}{\label{e7}}
 \frac{dp_{A\phi\phi}}{dt} &=& \{D_B(1-\rho_3^+)+D_A(1-\rho_0)\}p_{A\phi B}
          \nonumber\\&+&\epsilon(1-\rho_3^+)p_{AB\phi}
	  -(D_A+D_B)\rho_3^+p_{A\phi\phi},\nonumber\\
 \frac{dp_{A\phi B}}{dt} &=& (D_A+D_B)\rho_3^+p_{A\phi\phi}+\{D_B+D_A(1-
         \rho_0)\nonumber\\&+&\epsilon\rho_3^+\}p_{AB\phi}+D_A(1-\rho_0)
	 p_{ABB}-\{2D_A\nonumber\\&+&2D_B-D_A\rho_0-D_B\rho_3^+\}
	 p_{A\phi B},\nonumber\\
  \frac{dp_{AB\phi}}{dt} &=& \{D_B+D_A(1-\rho_3^+)\}p_{A\phi B}
         +\{D_B(1-\rho_3^+)\nonumber\\&+&\epsilon(1-\rho_3^+)\}p_{ABB}
         -\{D_A+D_B+\epsilon\nonumber\\&-&D_A\rho_0+D_B\rho_3^+\}
	 p_{AB\phi},\nonumber\\
  \frac{dp_{ABB}}{dt} &=& D_A\rho_3^+p_{A\phi B}+D_B\rho_3^+p_{AB\phi
        }-\{D_A(1-\rho_0)\nonumber\\&+&D_B(1-\rho_3^+)+\epsilon(1-
	\rho_3^+)\}p_{ABB}.
 \end{eqnarray}
 In order to solve the Eqs. (\ref{e7}) we need to know 
 $\rho_3^+$. To find the steady state solution of Eq. (\ref{e7}), and thus
 $\rho_1^+=p_{AB\phi}+p_{ABB}$, one could approximate $\rho_3^+=\rho_0$, 
 as was done for $l=2$ case. However, a better estimate for $\rho_3^+$ 
 can be obtained by writing a mean field equation for the densities
 $\rho_2^+$ and $\rho_3^+$ at second and third site ahead of the front,
 \begin{eqnarray}{\label{e8}}
 \frac{d\rho_2^+}{dt}&=&(\rho_1^++\rho_3^+)(1-\rho_2^+)D_B
         -D_B\rho_2^+(2-\rho_1^+-\rho_3^+),\nonumber\\
 \frac{d\rho_3^+}{dt}&=&(\rho_2^++\rho_4^+)(1-\rho_3^+)D_B
         -D_B\rho_3^+(2-\rho_1^+-\rho_3^+).\nonumber\\	 
 \end{eqnarray}
 In steady state Eqs. (\ref{e8}) gives, $\rho_3^+=(\rho_1^++2\rho_4^+)/3$.
 Now using the value of $\rho_1^+$ from Eq. (\ref{e5}) and approximating
 $\rho_4^+=\rho_0$, we obtain the following approximate expression for 
 $\rho_3^+$.
 \begin{eqnarray}{\label{e10}}
 \rho_3^+=\frac{3D_A\rho_0+3D_B\rho_0+2\epsilon\rho_0(1-\rho_0)}
         {3\{D_A+D_B+\epsilon(1-\rho_0)\}}
 \end{eqnarray}
 Using this value of $\rho_3^+$ in Eqs. (\ref{e7}), we obtain
 the steady state solution satisfying the normalisation $p_{A\phi\phi}
 +p_{A\phi B}+p_{AB\phi}+p_{ABB}=1$. Once we know these probabilities 
 the density at site just ahead of the front is obtained as: $\rho_1^+
 =p_{AB\phi}+p_{ABB}$. Substituting this value of $\rho_1^+$ in Eq. 
 (\ref{e2}), we obtained an improved analytic estimates for the 
 velocity as shown in the Fig. (\ref{fig:vel_2}). Here we 
 notice that as $l$ increases the data corresponding to the 
 analytic estimate comes closer to the simulation result. In this 
 approach one can systematically improve upon the estimates for 
 $\rho_1^+$ and thus $V$ by including states having successively 
 larger number of particles. We have also observed that
 when $D_A$=0, the front position $x$ grows with time $t$ as 
 $t^{1/2}$ as was reported in \cite{mai1}. Hence, when 
 $D_A=0$, the asymptotic velocity $V$ approaches to zero. Using Eq. 
 (\ref{e2}) for $D_A=0$, velocity is given as $V=\epsilon\rho_1^+$. The 
 condition for zero velocity simply implies that $\rho_1^+\rightarrow 0$.
 In the Fig. (\ref{fig:den_profile}), we have shown the simulation 
 results for the density profile ahead of the front for different 
 values of $D_A$ and notice that as $D_A\rightarrow$ 0, $\rho_1^+
 \rightarrow 0$, which gives $V\rightarrow 0$. We also note that when 
 $D_A=0$, $A$ particles form a compact cluster while $B$ particles 
 keep on moving diffusively. The number of $A$ particles created during 
 time interval $dt$, $dN_A=Vdt$ should be equal to the number of 
 consumed $B$ particles, $dN_B=V\rho_0dt$.
 However, the equality $dN_A=dN_B$ holds for non zero $V$ only if 
 $\rho_0=1$ while for $\rho_0\ne1$, it gives $V=0$. We also
 notice that when $D_A=0$, the front dynamics is like a DLA in $1d$,
 where, $A$ particles are static and growth occurs when $B$ particle
 diffuses and sticks with the neighbouring $A$ particle.
 \section{Conclusion}
   In this paper we have analyzed the reaction $A+B \rightarrow 2A$ 
   using an approximate analytic method and Monte Carlo simulation.
   Considering 
   the front as the position of rightmost $A$ particle we have been 
   able to obtain analytic estimate for its velocity 
   by writing master equation in the frame moving with the 
   front. One can improve upon the results systematically by 
   including states with successively larger number of sites 
   in front of the leading $A$ particle. The mean field equation
   (\ref{e1}) gives $V\propto \sqrt{D_A}$,
   but we observe linear dependence, $V\propto D_A$ for small $D_A$
   reflecting the important role of discrete fluctuations which is 
   neglected in the deterministic mean field equation (\ref{e1}). 
       

\begin{thebibliography}{}
  \bibitem{fisher} R. A. Fisher, Ann. Eugenics , 7, 355 (1937);
    A. Kolmogorov, I. Petrovsky and N. Piscounov, Moscow Uni. 
    Bull. Math. A1, 1 (1937).
  \bibitem{sar} W.van Saarloos, Phys. Rep. 386, 29 (2003).
  \bibitem{mai1} J. Mai, I. M. Sokolov, V. N. Kuzovkov and A. Blumen,
                Phys. Rev. E, 56, 4130 (1997).
  \bibitem{mai2} J. Mai, I. M. Sokolov and A. Blumen, Phys. Rev. Lett,
              77, 4462 (1996).
  \bibitem{mai3} J. Mai, I. M. Sokolov and A. Blumen, Europhys. Lett, 
            44, 7 (1998).
  \bibitem{sander} C. P. Warren, G. Mikus, E. Somfai and L. M. Sander,
             Phys. Rev. E, 63, 056103 (2001).
  \bibitem{smol} M. Smoluchowski, Z. Phys. Chem. (Munich) 92, 129 (1917);
                 S. O. Rice, Comprehensive Chemical Kinetics ( 
                 Elsevier, Amstardam, 1985 ), Vol. 25.
  \bibitem{derrida} E. Brunet and B. Derrida, Phys. Rev. E 56, 2597 (1997);
                   J. Stat .Phys. 103, 269 (2001).
  \bibitem{kessler} D. A. Kessler, Z. Ner and L. M. Sander, Phys. Rev. E 58, 107 (1998).
  \bibitem{ng} Niraj Kumar and G. Tripathy, cond-matt / 0505599.	
  \bibitem{kampen} N. G. van Kampen, Stochastic Processes in Physics
         and Chemistry (North-Holland, Amsterdam, 1981).
\end{thebibliography}
\end{document}